\documentclass[11pt]{article}
\usepackage{graphicx}
\begin{document}
\bibliographystyle{unsrt}
\def\ra{\rangle}
\def\la{\langle}
\def\beqn{\begin{equation}}
\def\eeqn{\end{equation}}
\def\bear{\begin{eqnarray}}
\def\eear{\end{eqnarray}}
\def\cdott{\cdot\cdot\cdot}
\def\bcen{\begin{center}}
\def\ecen{\end{center}}
\def\sx{SG_x}
\def\sy{SG_y}
\def\sz{SG_z}
\title{Stern-Gerlach Experiments and  Complex Numbers in Quantum Physics}
\author{S. Sivakumar\\Materials Physics Division\\ 
Indira Gandhi Centre for Atomic Research\\ Kalpakkam 603 102 INDIA\\
Email: siva@igcar.gov.in\\
}
\maketitle
\begin{abstract}
It is often stated that complex numbers are essential in quantum theory.  
In this article, the need for complex numbers in quantum theory is motivated using 
the results of tandem Stern-Gerlach experiments.      
\end{abstract}
\newpage
\section{Introduction}
 
       Complex numbers are essential in quantum theory.   In classical physics complex quantities are often introduced 
to  aid in solving problems rather than as a necessity.  That makes it mysterious for students about the role of complex numbers in quantum theory.     In this pedagogical report,  
it is illustrated that  the need for complex numbers in quantum theory can be 
made plausible after discussing the results of Stern-Gerlach (SG) experiment.  
This idea is presented in many texts, for instance, see the texts Sakurai\cite{sakurai}  
or Townsend\cite{townsend}.  Here, we wish to bring 
this to the notice of physics students and make a simplified presentation.\\

A SG apparatus is an arrangement to  provide spatially inhomogeneous magnetic 
field.  The purpose of spatial inhomogeneity  is to exert force on spins, which are 
like magnetic moments, so that spins of different orientations are spatially  
separated.  The direction of maximum gradient (a measure of inhomogeneity) is the 
axis along which spatial separation  of particles happens. If this direction is 
chosen to be the $z$-axis, the corresponding SG apparatus is said to be oriented 
along $z$-axis and it is denoted by $\sz$.  If the "spin" is indeed like a classical 
magnetic moment, then every possible orientation with respect to the orientation of 
the $\sz$ is possible.  So, the output beam is expected to be  continuously 
distributed along $z$ direction in  in space.  However, experiments indicated 
that there were finite number of output streams.   Particles in each of the stream is 
assigned a "spin" value.  If there are two outputs, the particles in one of the 
beams are said to be in up-spin state and those in the other output are said to be 
in the down-spin state.  Such particles are said to be "spin-half" particles.  Electrons, protons, neutrons, singly ionized silver atoms are some examples of 
spin-half particles.

\section{Recap of results of Stern-Gerlach experiment}
 
      The need for introducing complex numbers is easily 
recognized by knowing the results of experiments using two SG apparatuses in tandem.  
Consider a beam of spin-half system, for example,  singly ionized silver atoms,  
passing through a SG$_z$.  The output of the apparatus will have two 
beams that are spatially separated.  This indicates that the spin of the 
atoms in the beam  has two possible values.  In quantum 
theory this is taken to mean that the required state space is two-dimensional. 
Associated with these two possible spin values are two states, 
namely, $\vert z+\ra$ and $\vert z-\ra$.   
An arbitrary spin state $\vert\psi_{in}\ra$ is described by a superposition 
of the two states,
\beqn
\vert\psi_{in}\ra=r_1\vert z+\ra+r_2\vert z-\ra,
\eeqn
where $r_1$ and $r_2$ are the superposition coefficients that satisfy $r_1^2+r_2^2=1$.  
A short notation is used to present these facts.  A SG apparatus oriented along the $z$-axis is 
denoted by  $Z$ enclosed in a box.  The experimental fact that an arbitrary beam of 
spin-half systems  will give rise to two output beams is represented by
$$\vert\psi_{in}\ra\longrightarrow\framebox{Z}\longrightarrow\{\vert z+\ra,\vert z-\ra\},$$
where the states corresponding to the two output beams are enclosed in curly brackets.  
The relative intensities of the output beams decide the magnitude of the superposition 
coefficients.  Let us {\em assume that the superposition coefficients are real}.  According 
to the Born's rule for statistical interpretation, the relative 
intensities are  the squares of the respective superposition coefficients.  If the two output 
beams are of equal intensity, then the input state is a superposition of the two output states, 
\beqn
\vert\psi_{in}\ra=\frac{1}{\sqrt{2}}[\vert z+\ra+\vert z-\ra.
\eeqn

If the input beam is in the state $\vert z+\ra$, there is a single output beam corresponding. 
It is depicted by 
$$\vert z+\ra\longrightarrow\framebox{Z}\longrightarrow\vert z+\ra.$$
That is, $\vert z-\ra$ cannot be generated from $\vert z+\ra$ using $SG_z$.  Similarly, if the 
input state is $\vert z-\ra$, 
$$\vert z-\ra\longrightarrow\framebox{Z}\longrightarrow\vert z-\ra,$$
implying that $\vert z-\ra$ cannot be obtained from $\vert z+\ra$.   
In simple terms, $SG_z$ does not affect $\vert z+\ra$ and $\vert z-\ra$. Hence, they 
qualify as "eigenstates" of $SG_z$.  More importantly,  the fact that the state 
$\vert z+\ra$ cannot be generated from $\vert z-\ra$ and vice-versa, using $\sz$ 
implies that the two states $\vert z+\ra$ and $\vert z-\ra$ are "orthogonal" to each 
other.  In mathematical terms, orthogonality means the inner product between the two 
states vanishes.

   The choice of orientation of the SG apparatus is arbitrary. For instance, if the 
SG apparatus is oriented along $x$-direction, then an arbitrary input beam of spin-
half particles results in two output beams, separated spatially along the $x$-
direction.  The respective states of the particles in the two beams  
are denoted by $\vert x+\ra$ and $\vert x-\ra$.    As in the case of $SG_z$, the 
following are true:
$$\vert\psi_{in}\ra\longrightarrow\framebox{X}\longrightarrow\{\vert x+\ra,\vert x-\ra\},$$
$$\vert x+\ra\longrightarrow\framebox{X}\longrightarrow\vert x+\ra,$$
and
$$\vert x-\ra\longrightarrow\framebox{X}\longrightarrow\vert x-\ra.$$
And the conclusions are that the states $\vert x+\ra$ and $\vert x-\ra$ are 
orthogonal, eigenstates of $SG_x$.   Similarly, for the $y$-direction,
$$\vert\psi_{in}\ra\longrightarrow\framebox{Y}\longrightarrow\{\vert y+\ra,\vert y-\ra\},$$
$$\vert y+\ra\longrightarrow\framebox{Y}\longrightarrow\vert y+\ra,$$
and
$$\vert y-\ra\longrightarrow\framebox{Y}\longrightarrow\vert y-\ra.$$
As in the other cases,  the states $\vert y+\ra$ and $\vert y-\ra$ are orthogonal, 
eigenstates corresponding to $\sy$.

\subsection{Experiment I}

Are $\vert z+\ra$ and $\vert z-\ra$ unaffected by $\sx$?  To find out, 
one of the outputs of $\sz$, say, the beam of particles  corresponding to $\vert z+
\ra$, is used as input to $\sx$.  The experimental result is that there are two 
output beams of equal intensity.  So, from $\vert z+\ra$, both $\vert x+\ra$ and $\vert x-\ra$ emerge.  
Then the following assignment is  possible: 
\beqn
\vert z+\ra=\frac{1}{\sqrt{2}}[\vert x+\ra+\vert x-\ra],\label{IA}.
\eeqn
Once this choice is made for $\vert z+\ra$, the requirement for orthogonality 
implies that 
\beqn
\vert z-\ra=\frac{1}{\sqrt{2}}[\vert x+\ra-\vert x-\ra]\label{IB}.
\eeqn
These expressions are consistent with the requirement that $\vert z+\ra$ and $\vert z
-\ra$ are orthogonal to each other.  Note that the superposition coefficients are 
chosen to be real.  It does not matter if the expressions for the states 
$\vert z+\ra$ and $\vert z-\ra$ are swapped.

\subsection{Experiment II}
Let one of the outputs of $\sz$ be sent through a $\sy$.  Like the previous case, two 
output beams of equal intensity emerge from the apparatus.  Arguing as before, the results are 
\bear
\vert z+\ra&=&\frac{1}{\sqrt{2}}[\vert y+\ra+\vert y-\ra],\label{IIA}\\
\vert z-\ra&=&\frac{1}{\sqrt{2}}[\vert y+\ra-\vert y-\ra]\label{IIB},
\eear
where the superposition coefficients have been assumed to be real.  There is no inconsistency 
so far. 
\subsection{Experiment III}
The last piece of information required is to see the relation between the states 
$\vert x\pm\ra$ and $\vert y\pm\ra$.  For this, one of the output beams of $\sx$, 
for instance, the output  corresponding to $\vert x+\ra$,  is fed as  input to $\sy$.   Two 
output beams of equal intensity emerge. If the input is changed to $\vert x-\ra$, there 
are two output beams of equal intensity.  So, the results can be summarized as 
\bear
\vert x+\ra=\frac{1}{\sqrt{2}}[\vert y+\ra+\vert y-\ra],\\
\vert x-\ra=\frac{1}{\sqrt{2}}[\vert y+\ra-\vert y-\ra],
\eear
assuming that the superposition coefficients are real.
\section{Analysis of results}
What can be inferred from the results of the three experiments described above?  First of all, 
the conclusions of the  experiment III can be used to rewrite the results of the experiment II.  
This yields 
\bear
\vert z+\ra=\vert x+\ra,\\
\vert z-\ra=\vert x-\ra.
\eear
This is at variance with the results of experiment I which indicate that 
$\vert z+\ra$ and $\vert z-\ra$ are linear combinations of $\vert x+\ra$ and 
$\vert x-\ra$.  Obviously, one of the assumptions used in expressing the results 
should be wrong.  The crucial assumption made is that the input state is expressible 
as a linear combination of output states with {\em real} coefficients.  Now, it 
needs to be argued that using complex coefficients  yields consistent results.   The 
requirements are that the two output states are orthogonal and are of equal 
intensity.  So, one possibility is to recast the results of Experiment III using 
complex coefficients to give
\bear
\vert x+\ra=\frac{1}{2}[(1-i)\vert y+\ra+(1+i)\vert y-\ra],\label{IVA}\\
\vert x-\ra=\frac{1}{2}[(1+i)\vert y+\ra+(1-i)\vert y-\ra]\label{IVB}.
\eear
where $i=\sqrt{-1}.$  The definition of inner product between two states 
$\vert\psi_1\ra=a\vert z+\ra+b\vert z-\ra$ and 
$\vert\psi_2\ra=c\vert z+\ra+d\vert z-\ra$ is $\la\psi_1\vert\psi_2\ra=a^*c+b^*d$, 
where superposition coefficients $a,b,c$ and $d$ are complex numbers, and the 
superscript $*$ implies complex conjugation.    With this definition of inner 
product, the orthogonality condition is  satisfied.  Further, the coefficients are 
of equal magnitude to account for the observation that the output beams are of equal 
intensity.   This specific  choice of superposition coefficients ensures that the 
results of the Experiments I and II need not be rewritten with complex coefficients, 
and it concurs with the convention adopted in quantum physics.   Other choices such 
as 
\bear
\vert x+\ra=\frac{1}{\sqrt{2}}[\vert y+\ra+i\vert y-\ra],\label{IVC}\\
\vert x-\ra=\frac{1}{\sqrt{2}}[\vert y+\ra-i\vert y-\ra]\label{IVD},
\eear
would require rewriting the results of the experiments I and II with complex 
coefficients.  

\section{Discussion}
Complex numbers are essential in the Hilbert space formulation of quantum theory.     
Without invoking complex numbers, it is impossible to consistently explain the 
outcomes of some simple experiments performed with SG devices in tandem.  Another 
important point to note is that the Schrodinger equation has not been used in the 
arguments presented here.  Even though  $\sqrt{-1}$ appears explicitly in the 
Schrodinger equation which governs dynamics in quantum physics, the requirement for 
complex numbers is not due to this particular rule of dynamics.  It is the linear 
vector space structure that is crucial in necessitating  complex numbers in quantum 
theory.


\end{document}